\documentclass[apl,aps,twocolumn]{revtex4-1}

\usepackage{amsmath}
\usepackage{graphicx} 
\usepackage{dcolumn} 
\usepackage{bm} 
\usepackage{color} 
\usepackage{epstopdf}
\usepackage{amssymb}
\usepackage{verbatim}
\usepackage[colorlinks=true, pdfstartview=FitV, linkcolor=blue, citecolor=blue, urlcolor=blue]{hyperref} 
\usepackage{epstopdf}

\begin{document}

\def\1H        {{$^1$H\/ }}
\def\H        {{$^1$H\/ }}
\def\F        {{$^{19}$F\/ }}
\def\Hs       {{$^1$H\/}}
\def\Fs       {{$^{19}$F\/}}
\def\first    {{1$^{\rm st}$ \/}}
\def\second   {{2$^{\rm nd}$ \/}}
\def\third    {{3$^{\rm rd}$ \/}}
\def\th       {{$^{\rm th}$ \/}}
\def\eg       {{\it e.g.}}
\def\ie       {{\it i.e.}}
\def\etal     {{\it et al.}}

\newcommand{\addMD}[1]{\textcolor{magenta}{#1}}

\newcommand{\ee}[1]{\times 10^{#1}}
\newcommand{\mr}[1]{\mathrm{#1}}
\newcommand{\unit}[1]{\,\mathrm{#1}}
\newcommand{\um}{\,\mu{\rm m}}
\newcommand{\us}{\,\mu{\rm s}}
\newcommand{\uT}{\,\mu{\rm T}}
\newcommand{\kT}{k_{\rm B}T}
\newcommand{\kB}{k_{\rm B}}
\newcommand{\muB}{\mu_{\rm B}}
\newcommand{\rtHz}{\sqrt{\mr{Hz}}}
\newcommand{\degree}{^\circ}

\newcommand{\yn}{\gamma_n}
\newcommand{\tc}{t_c}
\newcommand{\fc}{f_c}
\newcommand{\fRF}{f_\mr{RF}}
\newcommand{\Dfdev}{\Delta f_\mr{dev}}
\newcommand{\fcenter}{f_\mr{center}}
\newcommand{\fL}{f_\mr{L}}
\newcommand{\Tp}{T_\mr{p}}
\newcommand{\tm}{\tau_\mr{m}}
\newcommand{\Qdamped}{Q_\mr{damped}}

\title{Accelerated nanoscale magnetic resonance imaging through phase multiplexing}%
\author{B. A. Moores$^{1}$}
\author{A. Eichler$^{1}$}
\email{eichlera@phys.ethz.ch}
\author{Y. Tao$^{1,2}$}
\author{H. Takahashi$^{1}$}
\author{P. Navaretti$^{1}$}
\author{C. L. Degen$^{1}$}
\affiliation{$^1$Department of Physics, ETH Zurich, Otto-Stern-Weg 1, 8093 Zurich, Switzerland}
\affiliation{$^2$Department of Chemistry, Massachusetts Institute of Technology, 77 Massachusetts Avenue, Cambridge, Massachusetts 02139, USA}
\date{\today}
\begin{abstract}
We report a method for accelerated nanoscale nuclear magnetic resonance imaging by detecting several signals in parallel. Our technique relies on phase multiplexing, where the signals from different nuclear spin ensembles are encoded in the phase of an ultrasensitive magnetic detector.  We demonstrate this technique by simultaneously acquiring statistically polarized spin signals from two different nuclear species (\Hs, \Fs) and from up to six spatial locations in a nanowire test sample using a magnetic resonance force microscope. We obtain one-dimensional imaging resolution better than $5$\,nm, and subnanometer positional accuracy.
\end{abstract}
\maketitle


Nanoscale magnetic resonance imaging (nanoMRI) is a promising, yet challenging microscopy technique for three-dimensional imaging of single objects with nanometer spatial resolution \cite{sidles91,degen08APL}. Among the advantages of nanoMRI are the possibility of site-specific image contrast, the absence of radiation damage, and the fact that only a single copy of an object is required. These qualities are particularly well-suited to provide structural information of large biomolecular complexes that are known to overwhelm nuclear magnetic resonance (NMR) spectroscopy and that evade crystallization for X-ray analysis. Recent proof-of-concept experiments showed that nanoMRI is capable of imaging individual virus particles in three dimensions with $<10\unit{nm}$ spatial resolution \cite{degen09}, as well as isotope-specific image contrast~\cite{mamin09}. The best detection sensitivities achieved to date are in the range of $10^1-10^4$ statistically polarized nuclear spins \cite{mamin13,staudacher13,loretz14,muller14}. NanoMRI has been demonstrated using several ultrasensitive signal detection techniques, especially magnetic resonance force microscopy (MRFM) \cite{sidles95,poggio10} and diamond-based magnetometry \cite{mamin13,staudacher13,schirhagl14}.

Although $\sim10$\,nm spatial resolution has been reached in several experiments \cite{degen09,mamin09,nichol13}, realizing this resolution in three-dimensional images required long averaging times.  For instance, imaging the proton density (\Hs) in a single tobacco mosaic virus required two weeks of data acquisition~\cite{degen09}, even for coarsely sampled data.  The long averaging times are prohibitive if one intends to refine voxel sizes or to image multiple nuclear spin species (e.g. \1H and $^{13}$C).  The slow data acquisition is in part due to the point-by-point measurement procedure where only a small subset of nuclei in a sample is detected at a given time.

An interesting avenue for speeding up the image acquisition process is to measure multiple signals in parallel and to use post processing to calculate the contributions from each individual signal.  Signal encoding is especially well-suited for MRI since nuclear spins can be separately addressed by radio-frequency (RF) pulses based on their differing Larmor frequencies.  In micron-to-millimeter scale MRI, Fourier-transform \cite{kumar75} and Hadamard \cite{bolinger88,eberhardt07hadamard} encoding provide efficient means for detecting the thermal (Boltzmann) polarization of nuclear spins.  

When imaging voxels are less than $\sim (100\unit{nm})^3$ the thermal polarization becomes exceedingly small and is dominated by statistical polarization fluctuations \cite{herzog14}.  It is the variance of these statistical fluctuations that then serves as the imaging signal \cite{degen07}.  Since variance measurements cannot be coherently averaged, traditional encoding techniques fail and parallel signal detection is considerably more challenging.  One effort for parallel detection of statistical spin polarization included the use of multiple detector frequencies \cite{oosterkamp10}. Unfortunately this approach is limited by detector bandwidth, and often only a single short-lived spin signal (which is the case for most biological samples) can be accommodated.  An exciting prospect is to Fourier encode statistically polarized nuclei by a correlation measurement \cite{kempf03, nichol13}, however these methods require pulsed gradients or mechanical shuttling and have yet to be applied to 3D imaging.

In this work we introduce a simple multiplexing technique capable of reducing data acquisition times for statistically polarized spins, irrespective of detector bandwidth or availability of pulsed gradients.  Our technique relies on phase multiplexing, whereby spin signals are encoded into the in-phase and quadrature channels of a phase-sensitive nuclear spin detector.  We demonstrate the technique by simultaneously acquiring spin signals from different nuclear species, and from multiple spatial locations in a nanowire test sample using MRFM.  Additionally, we show that the imaging resolution can be improved by subdividing voxels at undiminished signal-to-noise ratio (SNR).

In nanoscale MRI, the number of spins in an imaging voxel is measured as the variance $\sigma_M^2$ of the fluctuating nuclear magnetization $M(t)$ \cite{degen07}. Phase-sensitive detection can be obtained by periodically reversing the sign of $M(t)$ to create an oscillating magnetic signal, and by synchronizing this ac signal with the detector reference clock \cite{degen07,mamin13}. The signal is usually evaluated at $0^{\circ}$ and $90^{\circ}$ with a lock-in amplifier. However, when squaring the signal to calculate its variance, the signs of the two lock-in quadratures are lost and the phase cannot be evaluated. We overcome this problem by demodulating the signal at $0^{\circ}$, $45^{\circ}$, $90^{\circ}$, and $135^{\circ}$. This four-channel lock-in technique allows retrieving the signal phase from the variances of the four channels \cite{SM}. It thus reconciles nanoMRI with phase-sensitive detection in the regime of small, statistically polarized spin ensembles.

\begin{figure}
\includegraphics[width=\columnwidth]{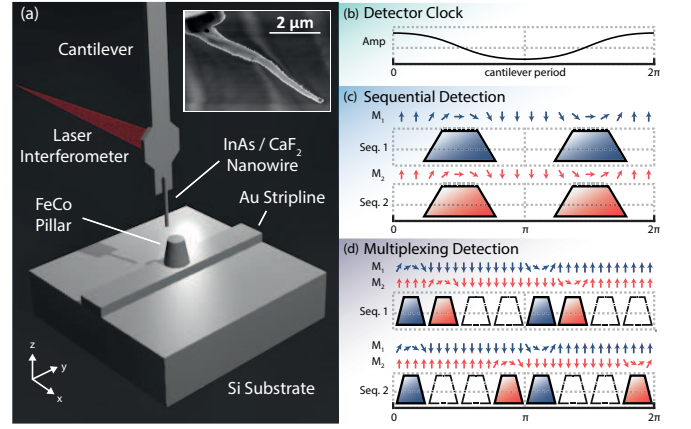}
\caption{Basic principle of phase multiplexing with $N=2$ nuclear spin ensembles.
(a) Schematic representation of the MRFM apparatus showing a micromechanical cantilever and ferromagnetic tip.
Inset: Scanning electron micrograph of an InAs nanowire test sample attached to the cantilever end. The nanowire is possibly terminated by a Au catalyst particle.
(b) Cantilever detector clock.  
(c) For sequential measurements, only a single nuclear spin ensemble is flipped at a time.  Arrows depict the orientation of nuclear magnetization $M_j(t)$ and trapezoids symbolize adiabatic RF pulses \cite{SM}.
(d) For phase multiplexed measurements, both nuclear spin ensembles are flipped, but the flipping is partially out of phase.  Two different flipping sequences are applied to generate two different measurements $E_1$ and $E_2$, which are subsequently reconstructed to yield $S_1$ and $S_2$.  Encoding phases in this example are $\phi_{11}=22.5\degree$, $\phi_{12}=67.5\degree$ (for $E_1$) and $\phi_{21}=22.5\degree$, $\phi_{22}=157.5\degree$ (for $E_2$).
}
\label{fig:model}
\end{figure}

In MRFM, the detector consists of a micromechanical cantilever that is coupled to the nuclear magnetization by the sharp magnetic gradient $G$ of a nanoscale ferromagnetic tip (see Fig. \ref{fig:model}(a)). The cantilever then converts the variance of the magnetic force $\sigma^2_\mr{F} = G^2 \times \sigma^2_\mr{M}$ into a measurable oscillating motion.

Phase multiplexing is achieved by exciting several nuclear spin ensembles simultaneously, while introducing a time delay between magnetization reversals.
Since signal detection is phase-sensitive, a time delay $\tau$ corresponds to a phase shift $\phi = 2\pi \tau/T$, where $T=1/\fc$ is the duration of one clock cycle and $\fc$ is the detector frequency. When measuring $N$ statistically independent spin ensembles, the total (complex) signal $E$ is

\begin{equation}
E = \sum_{j=1}^{N} S_j e^{i 2\phi_j} \ ,
\label{eq:total_signal}
\end{equation}
where $S_j=\sigma_{M_j}^2$ is the variance of the $j$'th ensemble's magnetization $M_j(t)$ and $\phi_j$ denotes the phase shift of the periodic reversal of $M_j(t)$ relative to the detector clock \cite{SM}.  In order to separate the different signal components, one may carry out $N$ measurements, each with a different combination of phases, yielding ${\vec{E}} = (E_1, E_2, ... , E_N)$. The reconstruction of the original spin signals ${\vec{S}} = (S_1, S_2, ... , S_N)$ then follows by linear recombination as
\begin{equation}
{\vec{S}} = \mathbf{A}^{-1} {\vec{E}} \ ,
\label{eq:reconstruction}
\end{equation}
where the transfer matrix
\[
\mathbf{A} =
\begin{bmatrix} 
e^{i 2\phi_{11}}  & e^{i 2\phi_{12}} & ...          & e^{i 2\phi_{1N}} \\ 
e^{i 2\phi_{21}}  & e^{i 2\phi_{22}} & ...          & e^{i 2\phi_{2N}} \\
\vdots 		       & \vdots                      & \ddots & \vdots  \\
e^{i 2\phi_{N1}} & e^{i 2\phi_{N2}} & ...  &  e^{i 2\phi_{NN}}
\end{bmatrix}
\]

contains the phase of each spin inversion during the cantilever period, as depicted in Fig. \ref{fig:model}(d). The phase at which the spin signal due to the $j$'th ensemble appears changes between different experiments $k$. Although any linearly independent set of phases $\phi_{jk}$ will allow for the reconstruction of ${\vec{S}}$, only a suitable choice of $\phi_{jk}$ will evade amplification of detector noise \cite{SM}.

Since each measurement $E_k$ simultaneously detects the magnetization of all nuclear ensembles, the signal collected after a complete sequence ${\vec{E}}$ is $N$ times larger compared to a sequential collection of $S_1, ... , S_N$ without multiplexing.  By contrast, the same amount of detector noise is added to each measurement $E_k$ regardless of whether multiplexing is applied.  Phase multiplexing can therefore improve the SNR by $\sqrt{N}$ for a fixed acquisition time.  Alternatively, the acquisition time can be reduced by $N$ without any loss in SNR.

Whether the improvement by $N$ is realized depends on the choice of phases $\phi_{jk}$.  Poorly selected phases will amplify detector noise when reconstructing ${\vec{S}}$ from ${\vec{E}}$.  We find that for white detector noise (such as thermal noise), the noise amplification factor is given by the matrix 2-norm: $||\mathbf{A}^{-1}||_2 = (\sum_{jk}|\tilde{a}_{jk}|^2)^{1/2} \geq 1$, where $\tilde{a}_{jk}$ are the matrix elements of $\mathbf{A}^{-1}$ \cite{SM}.  For an optimum set of phases $||\mathbf{A}^{-1}||_2=1$.  Although such an optimum set can be constructed (e.g., using the digital Fourier transform matrix $\phi_{jk} = \pi jk/N$), we used a heuristic search in order to satisfy additional constraints, especially peak RF pulse power.  Other potential noise sources include spin noise \cite{degen07}, correlations between spin ensembles, and instabilities in detector phase, gain or frequency \cite{SM}.

\begin{figure}[t]
\centering
\includegraphics[width=\columnwidth]{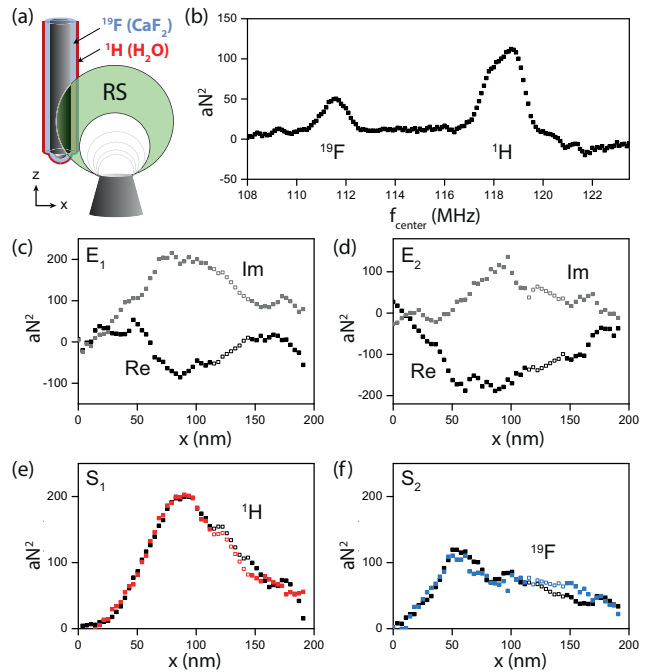}
\caption{(a) Schematic of the nanowire with \F and \1H layers.  RS represents the resonant slice, $\ie$ the volume in space where nuclear spins contribute to the signal.
(b) NMR spectrum of the sample measured by incrementing the RF center frequency $\fcenter$ while the nanowire was at a fixed position. In a magnetic field of $2.77\unit{T}$, the Larmor frequency of \1H is $118.0\unit{MHz}$ and of \F is $110.9\unit{MHz}$. Each data point was averaged for $360\unit{s}$.
(c,d) One-dimensional $x$ scans ($3.6\unit{nm}$ steps) at a fixed height using the multiplexing sequences of Fig. \ref{fig:model}(d).  Gray and black symbols represent the real and imaginary parts of $E_1$ and $E_2$, respectively. RF center frequencies were $118.0\unit{MHz}$ and $110.9\unit{MHz}$ with $\Dfdev = 0.5\unit{MHz}$.
(e,f) Reconstructed signal (colored squares) providing separate images for (e) \1H and (f) \Fs.
A sequential single isotope scan is shown for comparison (black squares).  Total averaging time per point was $240\unit{s}$ for multiplexed acquisition and $480\unit{s}$ for sequential acquisition. For reasons explained in \cite{SM}, the mechanical detector was unstable between $115$\,nm and $145$\,nm. Data in this range was replaced by interpolated data, represented by hollow symbols in (c-f). Data has been spatially low pass filtered by $5$ points.
}
\label{fig:chemical}
\end{figure}

We demonstrated phase multiplexing by measuring the statistical polarization of $\sim 10^5$ \1H and \F spins on an InAs nanowire test sample. The nanowire had a diameter of $120\unit{nm}$, and was coated with $60\unit{nm}$ of CaF$_2$ by thermal evaporation (see Fig. \ref{fig:chemical}(a)).  \1H spins were present in a $\sim 1\unit{nm}$ layer of surface adsorbates that naturally formed in ambient air \cite{degen09,mamin09,loretz14}.  For nanoscale MRI measurements, the nanowire was attached to the end of an ultrasensitive silicon cantilever and mounted in an MRFM apparatus operating at $4.2\unit{K}$ temperature and $2.77\unit{T}$ magnetic bias field.  Under measurement conditions, the cantilever had a resonant frequency $\fc \sim 5\unit{kHz}$, a spring constant of $k_c \sim 2.5 \times 10^{-4}$\,N/m, and a mechanical $Q\sim 30,000$, equivalent to a thermal force noise of about $3\unit{aN/\rtHz}$.  For the imaging, the nanowire was approached to within $100\unit{nm}$ of a $300$-nm-diameter FeCo magnetic tip \cite{poggio07}.

Nuclear magnetization reversals were performed using periodic application of adiabatic RF pulses.  These pulses had a center frequency $f_\mr{center}$ and bandwidth of $2\Dfdev$. They inverted nuclear spins only in a thin ``resonance slice'' (RS) in space whose Larmor frequencies $\fL = \yn B$ lay within $f_\mr{center}\pm\Dfdev$, where $B$ is the magnetic field at a spin's location (see Fig. \ref{fig:chemical}(a)), and $\yn$ is the nuclear gyromagnetic ratio.  For the multiplexing, several adiabatic pulses with different center frequencies $f_\mr{center}$ were co-added \cite{poggio07}.
Signal detection used a fiber-optic interferometer to read out the cantilever oscillation, and four-phase lock-in demodulation to extract the complex signal variance $E$ \cite{SM}. 

In a first experiment we performed multiplexing of two different nuclear isotopes (\1H and \Fs).  In order to identify the nuclear species, we parked the nanowire $\sim60\unit{nm}$ above the nanomagnetic tip and measured the nuclear magnetization as a function of RF center frequency (Fig. \ref{fig:chemical}(b)).  Two peaks at 111 MHz and 118 MHz confirmed the presence of both \1H and \F nuclear species.  To demonstrate multiplexing, we performed a one-dimensional spatial scan over the magnetic tip while exciting both \1H and \Fs.  At each location, $N=2$ signals were acquired with different phase sequences, resulting in the two scans $E_1$ and $E_2$ shown in Fig. \ref{fig:chemical}(c)-(d).  Application of Eq. (\ref{eq:reconstruction}) then directly reproduced the reconstructed signals (Fig. \ref{fig:chemical}(e)-(f)).  We found excellent agreement between reconstructed signals and sequential control measurements of \1H and \F.

\begin{figure}[t]
\centering
\includegraphics[width=\columnwidth]{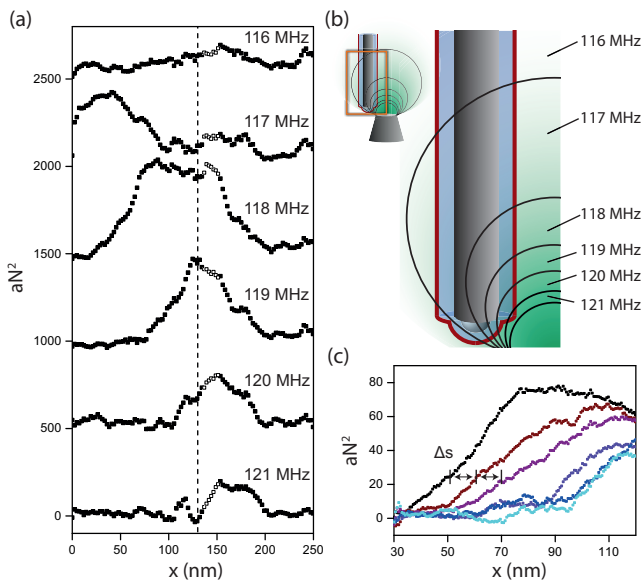}
\caption{(a) Signal from $N=6$ different $^1$H ensembles measured during a single $x$ scan with phase multiplexing. $x$ increment was $2.4\unit{nm}$, $\Dfdev = 0.3\unit{MHz}$, and center frequencies of resonant slices ranged from $116$ to $121\unit{MHz}$ as indicated. The averaging time was $360\unit{s}$ at each position for all $6$ measurements together. As in Fig. \ref{fig:chemical}, hollow points represent data that was interpolated due to instability of the mechanical oscillator \cite{SM}. Data was low pass filtered by $5$ points.
(b) Schematic of the spatial shape of resonant slices associated with Larmor frequencies $116-121\unit{MHz}$.  As the nanowire is scanned from left to right, spins intersect slices with progressively higher Larmor frequencies, reflected in a shifting peak in (a).  The schematic corresponds to the $x$ location of the vertical dotted line.
(c) High resolution $x$ scan with increments of $0.6\unit{nm}$, $\Dfdev = 0.13\unit{MHz}$, and slice center frequencies ranging from $117.8$ to $119.8\unit{MHz}$ in steps of $0.4\unit{MHz}$ (from left to right).  Averaging time was $1080\unit{s}$ at each position for all $6$ measurements together.  Data was low pass filtered by $30$ points.
}
\label{fig:space}
\end{figure}

In addition to chemical contrast imaging, multiplexing can also be applied to detect signals of the same isotope in different spatial regions of a sample. Such multi-slice imaging provides depth information with a single lateral scan, and may be useful to improve the fidelity of 3D image reconstruction.  Fig. \ref{fig:space}(a) shows an example of spatial multiplexing by detecting $N=6$ resonant slices of $^1$H.  A shifting peak is seen as the nanowire moves across the different slices, and the six signal traces clearly reflect the geometry of imaging slice and nanowire.  

We have tested multiplexing down to very low values of $\Dfdev$ to estimate the limits towards high spatial resolution. Figure \ref{fig:space}(c) shows such a high resolution scan acquired with frequency increments of $\Delta f_\mr{center} = 0.4\unit{MHz}$.  By comparing the signal onset as a function of $x$ position, we find that the lateral distance between slices $\Delta s$ is about $10\unit{nm}$.  This corresponds to a lateral magnetic gradient of $G = \partial B_z/\partial x = \Delta f_\mr{center}/(\yn \Delta s) \approx 1\ee{6}\unit{T/m}$.  A comparison with similar nanomagnetic tips, where $G\sim 4-5\ee{6}\unit{T/m}$ \cite{mamin12APL}, indicates that our tip had a lower-than-expected gradient, probably due to partial oxidation.  Note that the imaging resolution is not limited by the step size $\Delta s$, but by the bandwidth of the frequency modulation $\Dfdev=0.13\unit{MHz}$.  As the full width at half maximum of the resonant slice is approximately $\sqrt{2}\Dfdev$ \cite{degen09}, the imaging resolution is about $\sqrt{2}\Dfdev/(\yn G) \approx 4.3\unit{nm}$.  With an improved nanomagnetic tip, an imaging resolution of $\sim 1\unit{nm}$ can therefore be expected.

To compare the quality of multiplexed data to that of sequential measurements, we have quantified the signal error by analyzing the standard deviation of point-to-point fluctuations in the data sets.  For the multiplexed scans in Fig. \ref{fig:chemical}(e)-(f) we find $\epsilon_\mr{H} = 6.62\unit{aN^2}$ and $\epsilon_\mr{F} = 6.53\unit{aN^2}$, while the separate measurements have $\epsilon_\mr{H} = 6.89\unit{aN^2}$ and $\epsilon_\mr{F} = 4.94\unit{aN^2}$. The differences in amplitude between multiplexed and separately measured signals have standard deviations of $6.31$\,aN$^2$ for \1H and $8.34$\,aN$^2$ for \F \cite{SM}. These results confirm that phase multiplexing produces the same signal amplitude and SNR as sequential acquisition within half the averaging time.

While imaging step sizes and resolution for our experiments were on the order of a few nanometers, we found that the imaging precision was below one nanometer.  For example, in the leftmost trace of Fig. \ref{fig:space}(c) ($117.8$\,MHz center frequency), we observed that the signal rose with $3.4\unit{aN^2/nm}$ around $x\sim 65\unit{nm}$.  Comparing this slope with the measurement error $\epsilon_H = 0.8\unit{aN^2}$ we derive a position uncertainty of $0.24\unit{nm}$.  Although this precision is probably overestimated, we note that the signal onsets of the two scans in Fig. \ref{fig:chemical}(e) coincide within $0.6$\,nm \cite{SM}.  Such good positional accuracy is an important prerequisite for extending nanoMRI to subnanometer resolution.

Finally, we briefly comment on the limits of phase-multiplexed detection. When applying $N$ RF frequencies simultaneously, both average and peak power increase by at least $N$ regardless of the finer details of RF pulses.  Phase multiplexing therefore puts progressive demands on RF excitation, which in our experiments limited $N<10$.  Moreover, error analysis shows that strong signals tend to transmit noise to weak signals and eventually deteriorate the SNR of the latter \cite{SM}, which we found to become noticeable as $N>6$.

In summary, we have introduced a simple phase multiplexing method for accelerated detection of nanoscale NMR signals. The method is applicable even if spin ensembles are randomly polarized.  It can in principle be used with any phase-sensitive excitation/detection scheme, including those used in recent diamond-based magnetometry experiments \cite{mamin13,kotler11}. Using an MRFM apparatus, we have demonstrated simultaneous acquisition of nuclear spin signals from two different nuclear species and from up to six different sections within a sample. One-dimensional imaging scans reached a nominal spatial resolution $<5\unit{nm}$ with subnanometer positional accuracy.  The reduction in measurement time offered by our technique will be especially useful for 3D images of biomolecular complexes with isotope contrast.


\begin{acknowledgments}
We thank Cecil Barengo, Urs Grob, Christoph Keck, Heike Riel and the Physics machine shop for providing samples and experimental assistance.  We also appreciate the insightful discussions with Martino Poggio, Fei Xue, and Benedikt Herzog.  This work has been supported by the ERC through Starting Grant 309301, and by the Swiss NSF through the NCCR QSIT. B.A.M. acknowledges an NSERC fellowship. 
\end{acknowledgments}

\clearpage
\onecolumngrid

\large
\begin{center}

\textbf{Supplementary Material for: Accelerated nanoscale magnetic resonance imaging through phase multiplexing}

\normalsize

\vspace{5 mm}

Bradley A. Moores$^{1}$, Alexander Eichler$^{1}$, Ye Tao$^{1,2}$, Hiroki Takahashi$^{1}$, Paolo Navaretti$^{1}$, and Christian L. Degen$^{1}$

\textit{$^1$Department of Physics, ETH Zurich, Otto-Stern-Weg 1, 8093 Zurich, Switzerland, and}

\textit{$^2$Department of Chemistry, Massachusetts Institute of Technology, 77 Massachusetts Avenue, Cambridge, Massachusetts 02139, USA}

\end{center}

\small


\renewcommand{\thefigure}{S\arabic{figure}}
\renewcommand{\theequation}{S\arabic{equation}}
\renewcommand{\bibnumfmt}[1]{[S#1]}
\renewcommand{\citenumfont}[1]{S#1}
\setcounter{equation}{0}
\setcounter{figure}{0}
\renewcommand{\figurename}{\textbf{Supplementary Figure}}

\section{Measurement Setup}
\label{sec:measurementsetup}
Our magnetic resonance force microscope operated at a pressure of $<10^{-6}$\,mbar and at a temperature of $4.2$\,K in a liquid helium cryostat.  All components shown in Fig. 1 were mounted on a spring-suspended sample plate to attenuate mechanical vibrations.  The cantilever and laser interferometer were mechanically fixed, while the chip carrying the FeCo nanomagnet and the radio-frequency (RF) stripline could be mechanically scanned in three dimensions using remote controlled nanopositioners.  The laser for cantilever position readout had a wavelength of $1550$\,nm and an intensity of $100$\,nW, and was operated in a fiber-optic interferometer configuration formed between the end of a coated optical fiber and the cantilever.  The cantilever had a spring constant of $k_c \sim 2.5 \times 10^{-4}$\,N/m and a resonance frequency around $5$\,kHz.  The mechanical quality was around $Q=70,000$ at $4.2\unit{K}$, and dropped to about $30,000$ when brought close to the nanomagnet due to non-contact friction.  The cantilever was actively damped in a negative feedback loop using an FPGA controller in order to increase the measurement bandwidth.  The damped quality factor was on the order of $\Qdamped=400$, but would vary from $200-500$ depending on position as the magnetic tip was scanned over the cantilever.

\section{Adiabatic Spin Inversions}
\label{sec:adiabaticinversions}
Nuclear spins were inverted using adiabatic rapid passages based on frequency-modulated RF pulses. Two inversions were carried out within a cantilever period ($T = 0.2$\,ms) so that the nuclear magnetization became modulated at the resonance frequency $\fc=1/T$ of the cantilever. Two different frequency modulation schemes were used, including linear sweeps and sech/tanh-type sweeps.

For linear sweeps, the RF frequency was linearly increased from $f_\mr{center}-\Delta f_\mr{dev}$ to $f_\mr{center}+\Delta f_\mr{dev}$ over the duration of the pulse $\Tp$, which was equal to $T/(2p)$, where $p$ was the number of phase slots (see \ref{sec:transfermatrix}). The amplitude was modulated with a trapezoidal shape to ensure phase continuity at the beginning and end of each pulse.  Linear sweeps were used for the measurements shown in Fig. 2.

For sech/tanh-type sweeps, the frequency modulation followed a hyperbolic secant (\textit{sech}) and amplitude modulation used a hyperbolic tangent (\textit{tanh}) profile, which can improve adiabaticity~\cite{tomka13}.  In addition, consecutive pulses were allowed to have overlapping tails such that the pulse was effectively longer than $\Tp$, further aiding adiabaticity.  Sech/tanh sweeps (with $\beta=15$) were used for the measurements shown in Fig 3.

The maximum RF amplitude that could be employed in our apparatus was limited by undesired electrostatic driving (of the cantilever) that set in above $7\unit{mT}$, corresponding to a $300\unit{kHz}$ Rabi frequency for \H nuclei.  This allowed for about $p\sim 10$ phase slots before the adiabatic condition became violated.

\begin{figure}
\includegraphics[width=125mm]{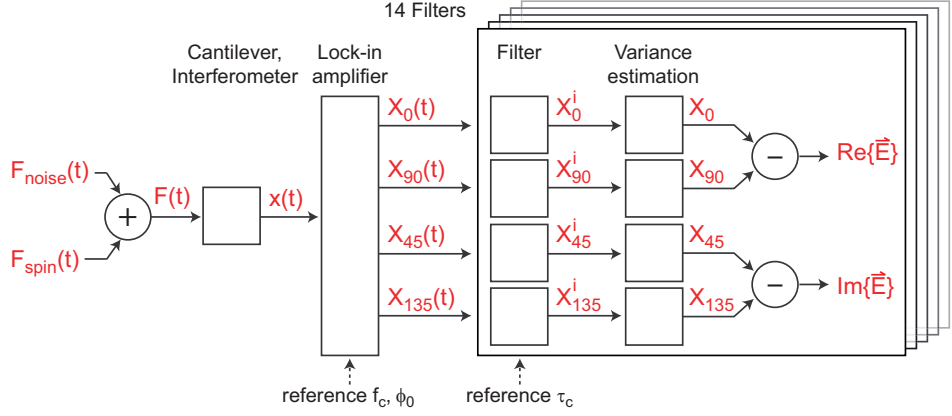}
\caption{\label{Figure S1} Schematic of signal acquisition protocol. $x(t)$ denotes the position of the cantilever as a function of time. $X_{0}(t)$, $X_{45}(t)$, $X_{90}(t)$, and $X_{135}(t)$ denote the cantilever motion amplitudes with different phases, which are then filtered with $14$ different time constants between $20$\,ms and $3$\,s. $X_{0}^i$, $X_{45}^i$, $X_{90}^i$, and $X_{135}^i$ are amplitudes filtered by the $i^\mr{th}$ filter. Finally, $X_{0}$, $X_{45}$, $X_{90}$, and $X_{135}$ stand for the variance of those amplitudes over the averaging time of the experiment (converted into units of squared force).}
\end{figure}

\section{Signal Acquisition and Reconstruction}
\label{sec:signalacquisition}
Nuclear spin noise is characterized by slow fluctuations of net magnetization, with time constants ranging between $\tm = 10-1000\unit{ms}$.  However, the cantilever motion is influenced by both undesired thermal noise ($F_\mr{noise}(t)$), and a much smaller force generated by the desired nuclear spin noise ($F_\mr{spin}(t)$).  These forces collectively influence the cantilever motion, which is then measured by the laser interferometer.

In order to detect the weak spin noise over the large thermal noise background, the cantilever motion can be measured using lock-in amplification and directing the spin signal into the in-phase channel.  The in-phase channel then receives both spin signal and thermal noise, while the quadrature channel only receives thermal noise.  If the variances of both channels are calculated and subtracted, one obtains a differential variance that now only contains the spin signal.  The signal-to-noise ratio can be optimized by choosing a lock-in filter time constant that is close to the intrinsic correlation time $\tm$.  We achieved this by calculating variances for a bank of filters and post-picking of the desired filter.

Since calculating the variance requires squaring the input signals, information concerning the sign (and hence the phase) of the signal is lost.  However, we note that the full phase of the spin signal can be retrieved by using a four-channel lock-in detection, where the interferometer signal is demodulated at the phases $0^{\circ}$, $45^{\circ}$, $90^{\circ}$ and $135^{\circ}$.  (A conventional lock-in amplifier would only demodulate at $0^{\circ}$ and $90^{\circ}$).  To understand this, consider measuring a spin signal with variance $\sigma^2_\mr{F}$ and phase $\phi$ in the presence of thermal noise with variance $\sigma^2_\mr{th}$.  If we denote the four demodulated signals by $X_{0}$, $X_{45}$, $X_{90}$, and $X_{135}$,
\begin{align}
&X_{0} = \sigma^2_\mr{F} \cos^2(\phi) + \sigma^2_\mr{th} \\
&X_{45} = \sigma^2_\mr{F} \cos^2(\phi - \pi/4) + \sigma^2_\mr{th} \\
&X_{90} = \sigma^2_\mr{F} \sin^2(\phi) + \sigma^2_\mr{th} \\
&X_{135} = \sigma^2_\mr{F} \sin^2(\phi - \pi/4) + \sigma^2_\mr{th}.
\end{align}
Note that the thermal noise, being incoherent, is equally distributed over the phase space.  We can subtract the thermal noise signal contributions by forming the differences
\begin{align}
X_{0} - X_{90} &= \sigma^2_\mr{F}(\cos^2(\phi) - \sin^2(\phi)) = \sigma^2_\mr{F} \cos(2\phi) \\
X_{45} - X_{135} &= \sigma^2_\mr{F}(\cos^2(\phi- \pi/4) - \sin^2(\phi- \pi/4)) = \sigma^2_\mr{F} \sin(2\phi).
\end{align}
These two terms correspond to the real and imaginary parts of a signal $E$,
\begin{equation}
E = (X_{0}-X_{90}) + i (X_{45} - X_{135}) = \sigma^2_\mr{F} e^{i2\phi}
\end{equation}
We implemented the four-channel lock-in detection technique by first demodulating by a conventional two-channel lock-in amplifier (SR830, Stanford Research), and then rotating lock-in outputs by $45^\circ$ on an FPGA controller.  The same FPGA then also performed the filtering and variance estimation to produce $X_{0}$, $X_{45}$, $X_{90}$, and $X_{135}$.  

For all imaging scans, two additional corrections were applied in order to account for position-dependent changes in cantilever gain (represented by the damped quality factor) and phase.  Both parameters were measured at each location prior to data collection.  For typical one dimensional scans, we found $\Qdamped$ to vary between $200-500$ while changes in phase were only within a few percent.  Since these changes vary slowly with position, roughly with a length scale similar to the tip-sample spacing, they can be easily corrected for by low-order polynomial fit or heavy low-pass filtering.  Note that these corrections apply to all imaging scans, including both single signal measurements and multiplexed measurements.

\section{Transfer Matrix}
\label{sec:transfermatrix}
Transfer matrices $\mathbf{A}$ were chosen such that the element-wise 2-norm of the inverse matrix was $||\mathbf{A}^{-1}||_2 \gtrsim 1$.  These matrices minimize the propagation of thermal noise, which is the dominant noise in our measurements (see \ref{sec:errorprop}).  To form the matrix, we first divided the cantilever half-period into $p$ equally spaced phase slots (\ie, $\phi_{jk} \in \{ 0^\circ, 30^\circ, 60^\circ, ... , 150^\circ \}$ for $p=6$).
We additionally required all phases within an experiment $E$ to be different in order to evenly distribute RF power over the cantilever period, thus minimizing peak power and spurious mechanical excitation. This constraint precludes the use of digital Fourier transform matrices, which would otherwise be the obvious choice.

In the case of multiplexing $N=2$ signals (Fig. 2), we used $p=4$ phase slots with matrix
\[
\mathbf{A} =
\begin{bmatrix} 
1  & e^{i \pi /2}  \\ 
1  & e^{i 3\pi/2}  \\
\end{bmatrix} \ ,
\]
This matrix has $||\mathbf{A}^{-1}||_2 = 1$.

In the case of multiplexing $N=6$ signals, we performed a heuristic search to find an optimal matrix.  The matrix used for Fig. 3 had $p=9$ with phases
\[
\mathbf{A} =
\begin{bmatrix} 
1  & e^{i 2\pi/3} & e^{i 14\pi/9} & e^{i 4\pi/3} & e^{i 2\pi/9} & e^{i 16\pi/9} \\ 
1  & e^{i 4\pi/3} & e^{i 8\pi/9} & e^{i 16\pi/9} & e^{i 10\pi/9} & e^{i 14\pi/9} \\
1  & e^{i 16\pi/9} & e^{i 2\pi/9} & e^{i 4\pi/3} & e^{i 14\pi/9} & e^{i 2\pi/3} \\
1  & e^{i 10\pi/9} & e^{i 8\pi/9} & e^{i 2\pi/3} & e^{i 2\pi/9} & e^{i 4\pi/9} \\
1  & e^{i 16\pi/9} & e^{i 2\pi/9} & e^{i 2\pi/3} & e^{i 4\pi/9} & e^{i 14\pi/9} \\
1  & e^{i 2\pi/3} & e^{i 14\pi/9} & e^{i 4\pi/9} & e^{i 10\pi/9} & e^{i 8\pi/9} \\
\end{bmatrix} \ .
\]
This matrix has $||\mathbf{A}^{-1}||_2 = 1.09$.  This means that the SNR improvement is reduced from $N=6$ to $N/||\mathbf{A}^{-1}||_2 = 5.5$.

\section{Error Propagation}
\label{sec:errorprop}
The reconstruction of phase multiplexed signals can sometimes lead to enhanced noise compared to single sweeps, which in turn deteriorates multiplexing performance. We have analyzed the effect of detector noise, spin noise, detector gain, and phase fluctuations on the reconstructed signal. In our experiments, detector noise was typically the dominating noise. 

\subsection{Detector Noise}
In MRFM, detector noise arises from thermal force noise acting on the micromechanical cantilever. It leads to thermomechanical noise with a Lorentzian distribution as shown in Fig.~\ref{Figure S2} (b), and is added equally to all measurements. If the thermal noise leads to a standard error $\epsilon_\mr{thermal}$ in each experiment $E_k$ (on both the real and imaginary channel), error propagation predicts a standard error on reconstructed signals $S_\mr{j}$ as
\begin{equation}
\epsilon_{S,j}^2 = \sum_{k=1}^{N} |\tilde{a}_{jk}|^2 \epsilon_\mr{thermal}^2 =  \epsilon_\mr{thermal}^2 \sum_{k=1}^{N} |\tilde{a}_{jk}|^2
\end{equation}
where $\tilde{a}_{jk}$ are the matrix elements of the inverse transfer matrix ${\bf A}^{-1}$.  A ``good'' transfer matrix will have little amplification of $\epsilon_\mr{thermal}^2$. 
As a measure for the total propagated error we use the sum of all individual errors $\epsilon_{S,j}^2$
\begin{equation}
\epsilon_S^2 = \sum_{j=1}^{N} \epsilon_{S,j}^2 = \sum_{j,k=1}^{N} |\tilde{a}_{jk}|^2 \epsilon_\mr{thermal}^2 = ||{\bf A}^{-1}||_2^2 \epsilon_\mr{thermal}^2
\end{equation}
Here $||...||_2$ denotes the element-wise 2-norm of the matrix.  As shown above, one can often find transfer matrix phases such that $||{\bf A}^{-1}||_2=1$ or $||{\bf A}^{-1}||_2\gtrsim 1$. Thus, amplification of thermal noise can be avoided by a suitable choice of ${\bf A}$.

\subsection{Spin noise}

Spin noise reflects the uncertainty in the variance measurement of spin fluctuations.  If the spin signal has variance $\sigma^2_\mr{F}$, the standard error of this variance estimate is
$\epsilon_\mr{spin} \approx (2/n)^{1/2} \sigma^2_\mr{F}$, where $n\approx T_{total}/\tau_m$ is the number of independent samples that can be performed during measurement time $T_{total}$ and $\tau_m$ is the correlation time of nuclear spin fluctuations~\cite{degen07sup}.

In a multiplexed experiment, each spin signal $\sigma^2_\mr{F,j}$ carries a spin noise error $\epsilon_{\mr{spin},j}$.  This error is propagated to the measurements $E_k$ as
\begin{equation}
\epsilon_{E,k}^2 = \sum_{j=1}^{N} \left\{ \Re[a_{jk}]^2 \epsilon_{\mr{spin},j}^2 + i \Im[a_{jk}]^2 \epsilon_{\mr{spin},j}^2 \right\}
\end{equation}
and to the reconstructed signals $S_j$ as
\begin{eqnarray}
\epsilon_{S,j}^2 & = & \sum_{j,k=1}^{N} \left\{ \Re[\tilde{a}_{jk}]^2 \Re[a_{jk}]^2 \epsilon_{\mr{spin},j}^2 + \Im[\tilde{a}_{jk}]^2 \Im[a_{jk}]^2 \epsilon_{\mr{spin},j}^2 \right\} \\
                 & = & \sum_{j=1}^{N} \epsilon_{\mr{spin},j}^2 \sum_{k=1}^{N} \left\{ \Re[\tilde{a}_{jk}]^2 \Re[a_{jk}]^2 + \Im[\tilde{a}_{jk}]^2 \Im[a_{jk}]^2 \right\}
\end{eqnarray}
$\Re[...]$ and $\Im[...]$ denote real and imaginary parts. The expression above allows one to compute explicitly the error introduced by spin noise for a specific choice of transfer matrix {\bf A}.  

If one does not want to calculate errors explicitly, as an approximation, one finds that
\begin{equation}
\sum_{k=1}^{N} \left\{ \Re[\tilde{a}_{jk}]^2 \Re[a_{jk}]^2 + \Im[\tilde{a}_{jk}]^2 \Im[a_{jk}]^2 \right\} \approx \frac{1}{2N}
\end{equation}
This follows from $||{\bf A}||_2 = N^2$ and  $||{\bf A}^{-1}||_2 \approx 1$ and assuming that, approximately, all entries have real and imaginary parts of similar magnitudes.  The error added to each reconstructed signal is therefore approximately
\begin{equation}
\epsilon_{S,j}^2 \approx \frac{1}{2N} \sum_{j=1}^{N} \epsilon_{\mr{spin},j}^2
\end{equation}
That is, the strongest spin noise will transmit to all reconstructed signals $S_j$, and all reconstructed signals will now carry a similar error.  This may be an issue if one tries to detect a weak signal in the presence of a strong signal.

\subsection{Detector phase variation between experiments}

If there are variations in the detector reference phase between experiments $E_k$, reconstruction will lead to cross-talk between the $S_j$ that appears as additional noise. In MRFM, detector phase fluctuations are due to changes in the cantilever resonance frequency that alter the phase response. Since phase fluctuations affect all signals equally, no relative phase errors are expected between different signals within an experiment $E_k$.

In a multiplexed experiment, phase variations will mainly occur between different experiments $E_k$. If we assume that there is a small phase uncertainty $\delta\phi_k \ll 1$ between different experiments $k$, the matrix elements $a_{jk}$ become
\begin{equation}
a'_{jk} = e^{i(\phi_{jk}+\delta\phi_k)} = e^{i\phi_{jk}} e^{i\delta\phi_k} \approx a_{jk} (1+i\delta\phi_k)
\end{equation}
If $\epsilon_\phi$ denotes the standard deviation of $\delta \phi_k$, then the error propagates to the measurements $E_k$ as
\begin{equation}
\epsilon_{E,k}^2 = \sum_{j=1}^{N} (\sigma^2_\mr{F,j})^2 \epsilon_{\phi}^2 \left\{ \Im[a_{jk}]^2  + i \Re[a_{jk}]^2 \right\} 
\label{eq:phase_noise1}
\end{equation}
and to the reconstructed signals $S_j$ as
\begin{eqnarray}
\epsilon_{S,j}^2 & = & \sum_{j,k=1}^{N} (\sigma^2_\mr{F,j})^2 \epsilon_{\phi}^2 \left\{ \Re[\tilde{a}_{jk}]^2 \Im[a_{jk}]^2 + \Im[\tilde{a}_{jk}]^2 \Re[a_{jk}]^2 \right\} \\
                 & = & \sum_{j=1}^{N} (\sigma^2_\mr{F,j})^2 \epsilon_{\phi}^2 \sum_{k=1}^{N} \left\{ \Re[\tilde{a}_{jk}]^2 \Im[a_{jk}]^2 + \Im[\tilde{a}_{jk}]^2 \Re[a_{jk}]^2 \right\}
\end{eqnarray}
From the above expression one finds again that, approximately, the error added to each reconstructed signal is
\begin{equation}
\epsilon_{S,j}^2 \approx \frac{1}{2N} \epsilon_{\phi}^2 \sum_{j=1}^{N} (\sigma^2_\mr{F,j})^2
\label{eq:phase_noise4}
\end{equation}
That is, phase noise again transmits to all reconstructed signals and the dominating effect will come from the strongest signal.  As with spin noise, this may be an issue if one tries to detect a weak signal in the presence of a strong signal.

\subsection{Detector phase variation during an experiment}

Fast fluctuations in the cantilever resonance frequency (jitter) will change both the phase and amplitude response of the detector. Since the amplitude is only affected to second order in frequency, the dominant influence is through the phase response.

If we consider a small frequency jump $\delta f_c < f_c/\Qdamped$ during an experiment, with $f_c$ the cantilever resonance frequency and $\Qdamped$ the feedback-damped quality factor, the corresponding phase jump will be $\delta \phi \approx -\Qdamped \delta f_c/f_c$.  If $\epsilon_{f_c}$ is the standard deviation of frequency jumps during an experiment, then the associated standard deviation of phase is
\begin{equation}
\epsilon_\phi \approx \frac{\Qdamped }{f_c} \epsilon_{f_c}
\end{equation}
and it will affect the reconstructed signal $S_j$ in the same way as given in equations \ref{eq:phase_noise1} through \ref{eq:phase_noise4}.

\subsection{Detector gain variation}

If there are variations in the detector gain between experiments $E_k$, reconstruction will likewise lead to cross-talk between the $S_j$.  In MRFM, detector gain fluctuations are due to changes in the cantilever resonance frequency that alter the amplitude response, or due to variations in the damped quality factor $Q_{damped}$.

As with phase variations, gain variations will occur between different experiments $E_k$, but no relative gain errors are expected between different signals within an experiment $E_k$.  If we assume that there is relative gain uncertainty $\delta g_k \ll 1$ between different experiments $k$, the matrix elements $a_{jk}$ become
\begin{equation}
a'_{jk} = (1+\delta g_k) e^{i\phi_{jk}} = a_{jk} (1+\delta g_k)
\end{equation}
If $\epsilon_g$ denotes the standard deviation of $\delta g_k$, then the error propagates to the measurements $E_k$ as
\begin{equation}
\epsilon_{E,k}^2 = \sum_{j=1}^{N} (\sigma^2_\mr{F,j})^2 \epsilon_{g}^2 \left\{ \Re[a_{jk}]^2  + i \Im[a_{jk}]^2 \right\} 
\end{equation}
and to the reconstructed signals $S_j$ as
\begin{eqnarray}
\epsilon_{S,j}^2 & = & \sum_{j,k=1}^{N} (\sigma^2_\mr{F,j})^2 \epsilon_{g}^2 \left\{ \Re[\tilde{a}_{jk}]^2 \Re[a_{jk}]^2 + \Im[\tilde{a}_{jk}]^2 \Im[a_{jk}]^2 \right\} \\
                 & = & \sum_{j=1}^{N} (\sigma^2_\mr{F,j})^2 \epsilon_{g}^2 \sum_{k=1}^{N} \left\{ \Re[\tilde{a}_{jk}]^2 \Re[a_{jk}]^2 + \Im[\tilde{a}_{jk}]^2 \Im[a_{jk}]^2 \right\}
\end{eqnarray}
From the above expression one finds again that, approximately, the error added to each reconstructed signal is
\begin{equation}
\epsilon_{S,j}^2 \approx \frac{1}{2N} \epsilon_{g}^2 \sum_{j=1}^{N} (\sigma^2_\mr{F,j})^2
\end{equation}

\section{Comparison of noise between single and multiplexed scans}

We used three methods to analyze the noise affecting our signal. First, the \textbf{point-to-point fluctuations} of a data set can be quantified by the standard deviation $\epsilon_\mr{pp}$ of the data points, or, equivalently, by the standard deviation of the \textit{differences} of consecutive points $\epsilon_\mr{diff}$. According to error propagation, $\epsilon_\mr{diff} = \epsilon_\mr{pp} \sqrt{2}$. In practice, we found that $\epsilon_\mr{diff}$ is a more reliable measure of the noise than $\epsilon_\mr{pp}$ because it is less susceptible to a signal bias that will artificially increase $\epsilon_\mr{pp}$. We therefore calculated $\epsilon_\mr{diff} / \sqrt{2}$ for all sweeps and used it as an upper bound for the real signal fluctuations $\epsilon_\mr{H}$ or $\epsilon_\mr{F}$.

Second, for each data point, we calculated the \textbf{statistical uncertainty} $\epsilon_\mr{stat}$ of the signal (which should be equal to $\epsilon_\mr{pp}$). For a statistical quantity with mean $Z$ that is sampled $n$ times, this amounts to $\epsilon_\mr{stat}^2 = 2Z/n$~\cite{degen07sup}. In our case, $Z$ corresponds to the signal $S_\mr{j}$ while $n$ is the number of statistically independent samples during an averaging period.

Third, we use a \textbf{noise model} to compute the signal fluctuations as $\epsilon_\mr{model}$. The model includes thermal vibrations of the cantilever, amplifier noise, and laser shot noise. From this model, we can estimate the contributions of the different noise sources. We take into account the effective temperature of the mechanical mode, which is lower than the cryostat temperature since the cantilever is actively cooled by a feedback loop. However, $\epsilon_\mr{model}$ will slightly underestimate the total signal fluctuations because it does not include the spin noise, i.e. it neglects the fact that our signal is itself the variance of a fluctuating quantity.

\begin{figure}
\includegraphics[width=125mm]{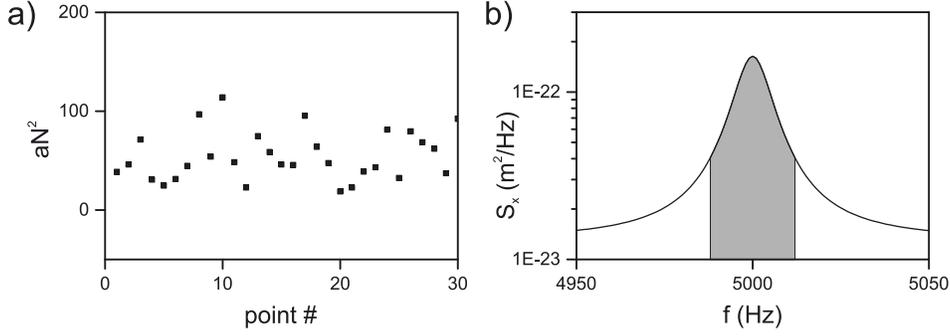}
\caption{\label{Figure S2} \textbf{(a)} Data points measured as a function of time at a fixed position. Averaging was performed over $60$\,s with a slice center frequency of $117.8$\,MHz and $\Delta f_\mr{dev} = 130$\,kHz. \textbf{(b)} Power spectral density of the apparent cantilever displacement noise calculated with a model that simulates the parameters used in \textbf{(a)}. The noise stems from thermal vibrations, amplifier noise, and laser shot noise. The number of statistically independent samples at each point is $1936$, and the effective filter width of our $1^\mr{st}$ order Butterworth filter is $\Delta f_\mr{filter} = 12.4$\,Hz (marked by a shaded region). The damped mode temperature is estimated to be $59$\,mK.}
\end{figure}

We compared the values obtained with all three methods. For this comparison, we measured a signal $30$ times at the same location to avoid calibration uncertainties arising from varying surface interaction as a function of $x$. Figure~\ref{Figure S2} (a) shows the absolute value of the measured signal. We obtained $\epsilon_\mr{pp} = 24.9$\,aN$^2$, $\epsilon_\mr{diff} / \sqrt(2) = 24.1$\,aN$^2$, $\epsilon_\mr{stat} = 26.0$\,aN$^2$, and $\epsilon_\mr{model} = 23.2$\,aN$^2$. The modeled noise spectral density $S_\mr{x}$ is shown in Fig.~\ref{Figure S2} (b). As mentioned above, $\epsilon_\mr{model}$ is slightly smaller than the other two values because it neglects the influence of spin noise. The contribution of the thermomechanical noise amounted to $20.1$\,aN$^2$ and that of the electrical amplifier noise to $3.1$\,aN$^2$. The laser shot noise was negligible with a contribution of $0.01$\,aN$^2$. Based on the good agreement between different noise calculations, we relied on $\epsilon_\mr{diff}/\sqrt{2}$ to quantify the noise because it allowed a direct measure of signal fluctuations.

As a measure for the overall fidelity of multiplexed signals relative to separately measured signals, we subtracted the two data curves for \H in Fig. 2e from each other and evaluated the standard deviation of the point-to-point fluctuations $\epsilon_\mr{sub}$ of the result (and likewise for the two data curves for \F in Fig. 2f). From error propagation, the difference of two data sets is expected to have larger noise than an individual data set by $\sqrt{2}$. We found $\epsilon_\mr{sub}/\sqrt{2} = 6.31$\,aN$^2$ for \H and $8.34$\,aN$^2$ for \F, which is close to the values we obtained for all four individual curves (see main text).

\section{Estimation of positional accuracy}

The point-to-point fluctuations that arise mainly from thermomechanical noise and amplifier noise (see previous section) are not only a measure of the signal quality, but also set the resolvable scanning step size. A step in $x$ with magnitude $\Delta x$ is resolvable when the signal change exceeds the signal noise, i.e. when  $\Delta x \times d S / d x$ is at least as large as the standard deviation of the point-to-point fluctuations. The filtered data in Fig. 2b of the main text have fluctuations with standard deviations of $\epsilon_H = 6.62$\,aN$^2$ and $\epsilon_H = 6.89$\,aN$^2$ for multiplexed and single sweep data, respectively. Compared with the maximum signal slope of $\sim 6.2$\,aN$^2/$nm around $60$\,nm, this gives a best positional accuracy of close to $1$\,nm. The data in Fig. 3a with center frequencies $118$\,MHz and $119$\,MHz have fluctuations with standard deviations equal to $\epsilon_H = 15$\,aN$^2$ and $\epsilon_H = 13.8$\,aN$^2$, maximum signal slopes of $25.0$\,aN$^2/$nm and $25.4$\,aN$^2/$nm around $65$ and $120$\,nm, and thus best positional accuracies corresponding to $0.6$\,nm and $0.54$\,nm, respectively. The data in Fig. 3d with center frequency $117.8$\,MHz have fluctuations with a standard deviation of $0.827$\,aN$^2$, a maximum signal slope of $3.38$\,aN$^2/$nm around $65$\,nm, and thus a best positional accuracy of $0.24$\,nm. The fluctuation standard deviations vary between different data sets because of different averaging times, different low-pass filtering, and because the effective temperature of the cantilever mechanical mode depends on the undamped and damped mechanical quality factors.

\begin{figure}
\includegraphics[width=80mm]{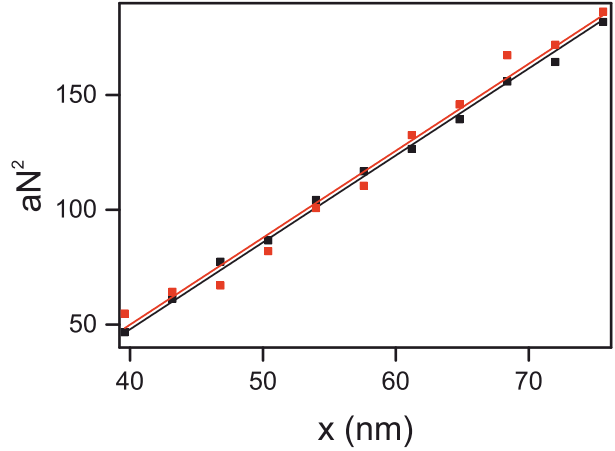}
\caption{\label{Figure S3} Linear fits to multiplexed and separately measured data for \H. The plot is a zoom of Fig. 2e of the main text. The two linear fits were performed using a common slope. The resulting x-offset is $0.58$\,nm.}
\end{figure}

An alternative estimation of the positional accuracy can be obtained by comparing the signal onset for the multiplexed and separately measured \H signals in Fig. 2e of the main text. Figure~\ref{Figure S3} shows a zoom of the two data sets. We first performed a linear fit to both data sets to determine a common slope ($3.79$\,aN$^2/$nm). We then performed linear fits to the two sets individually where we kept the slope fixed. The resulting offset in $x$ between the fitted lines is $0.58$\,nm.

\section{Mechanical coupling between cantilevers}

X-position scans presented in this work contain data points with hollow squares that correspond to spin signals with higher uncertainty. At these positions, we observed that the cantilever was driven to a large amplitude in spite of the damping feedback. This made a ringdown measurement, and therefore a precise determination of the quality factor, impossible. This problem arose whenever the resonance frequency of the measured cantilever matched that of a second cantilever mounted on the same chip ($\sim 5051$\,Hz). We believe that the feedback loop is then responsible for a coupling that drove the measured cantilever. Interpolated values were used to approximate the spin signals at these locations.

\end{document}